# Detection of virulence factors and β lactamase encoding genes among the clinical isolates of *Pseudomonas aeruginosa.*

Fazlul MKK[1], Najnin A[2], Farzana Y[3], Rashid MA[4], Deepthi S[5], Srikumar C[5], SS Rashid[1], Nazmul MHM[5]

[1]1Faculty of Industrial Sciences Technology, Universiti Malaysia Pahang, Gambang, 26300 Pahang, Malaysia

[2]Jeffrey Cheah School of Medicine and Health Sciences, Monash University, No.8, Jalan Masjid Abu Bakar, 80100 Johor Bahru, Malaysia

[3]Faculty of Science, Lincoln University, 12-18, Jalan SS6/12, Off Jalan Perbandaran, 47301 Petaling Jaya, Selangor Malaysia.

[4]Faculty of Medicine, University Teknologi MARA, Jalan Hospital, Sg Buloh, Selangor 47000, Malaysia

[5]Center of Research Excellence, Graduate School of Medicine, Perdana University, Jalan MAEPS Perdana, Serdang, 43400 Selangor, Malaysia

## Abstract

**Background:** Pseudomonas aeruginosa has emerged as a significant opportunistic bacterial pathogen that causes nosocomial infections in healthcare settings resulting in treatment failure throughout the world. This study was carried out to compare the relatedness between virulence characteristics and β-lactamase encoding genes producing Pseudomonas aeruginosa.

**Methods:** A total of 120 P. aeruginosa isolates were obtained from both paediatric and adult patients of Selayang Hospital, Kuala Lumpur, Malaysia. Phenotypic methods were used to detect various virulence factors (Phospholipase, Hemolysin, Gelatinase, DNAse, and Biofilm). All the isolates were evaluated for production of extended spectrum beta-lactamase (ESBL) as well as metallo β-lactamase (MBL) by Double-disk synergy test (DDST) and E-test while AmpC β-lactamase production was detected by disk antagonism test.

**Results:** In this study, 120 Pseudomonas aeruginosa isolates (20 each from blood, wounds, respiratory secretions, stools, urine, and sputum samples) were studied. Among Pseudomonas aeruginosa isolates, the distribution of virulence factors was positive for hemolysin (48.33%), DNAse (43.33%), phospholipase (40.83%), gelatinase (31.66%) production and biofilm formation (34%) respectively. The prevalence of multiple β-lactamase in P. aeruginosa showed 19.16% ESBL, 7.5% MBL and 10.83% AmpC production respectively.

**Conclusion:** A regular surveillance is required to reduce public health hazard and the spread of virulence factors and β-lactamase genes among clinical isolates of Pseudomonas aeruginosa

## 1. INTRODUCTION

The widespread expansion of *Pseudomonas aeruginosa* resistant strains is due to exploitation and mistreatment of antibiotics which generates several bacterial virulences.

Antibiotics play a vital role against pathogenic bacteria, but studies have shown that antibiotics to be less effective against bacteria with different virulence factors. The improvement and innovation of antimicrobials agents with novel or unexplored properties are becoming an increasing need to control and manage bacterial infectious diseases effectively. *Pseudomonas aeruginosa* is opportunistic pathogenic bacteria, causes numerous infection. *P. aeruginosa* releases various types of virulence factors, and the pathogenesis is linked with several cell-associated and extracellular factors which are responsible for the severe therapeutic problem (Alireza Mohammadzadeh, Jalal Mardaneh, Reza Ahmadi, & Javad Adabi, 2017). Infectious diseases caused by *P. aeruginosa* are problematic to treat due to the resistance development and produce multiple virulence factors (Pereira, Rosa, & Cardoso, 2015). In the hospital, commonly caused nosocomial infection particularly in immunosuppressed, burns and cystic fibrosis patients are due to *Pseudomonas aeruginosa*. Severe infections with high mortality are problematic to treat and often failure treatment. The high percentage of mortality and its multidrug resistance mechanism has driven *P. aeruginosa* as an emerging superbug (Khan & Khan, 2016; Tanwar, Das, Fatima, & Hameed, 2014). Virulence factors modify the immune system by adhesion, evading and destroying the tissues for the progression of diseases (Moghaddam, Khodi, & Mirhosseini, 2014).

*P. aeruginosa* possesses many virulence factors (A. Mohammadzadeh, J. Mardaneh, R. Ahmadi, & J. Adabi, 2017; Sánchez-Diener et al., 2017; Ullah, Qasim, Rahman, Jie, & Muhammad, 2017). *Pseudomonas aeruginosa* is widely distributed and able to secrete multiple virulence factors such as hemolysin, gelatinase, DNAase and produce biofilm (Pramodhini, Umadevi, & Seetha, 2016). These factors are damaging their host's immune systems and form a barrier to antibiotics that reduce antibiotic efficiency resulting in treatments to be incompetent and failure. Numerous virulence factors of *P. aeruginosa* can cause acute and chronic infections (Sánchez-Diener et al., 2017). Several extracellular virulence factors are liable for blood steam invasion, diffusion, and tissue damage after colonization (Sonbol, Khalil, Mohamed, & Ali, 2015). Hemolysin and protease are known to be important virulence factors of *P. aeruginosa*. The formation of hemolysin production in *Pseudomonas aeruginosa* is dependent on the isolates specimens and antigenic structure (Kamel, Edeen, Yousef El-Mishad, & Ezzat, 2011). Gelatinase is a group of protease which can hydrolyze gelatine into a polypeptide, amino acid, and peptides (Balan, Nethaji, Sankar, & Jayalakshmi, 2012). DNase degrades the extracellular DNA to be weak and susceptible to microorganisms (Sharma & Pagedar Singh, 2018). Biofilm formation in *P. aeruginosa* reduces the efficacy of antibiotic sensitivity, and metabolic rate (Mulcahy, Isabella, & Lewis, 2014) but enhance efflux (Alav, Sutton, & Rahman, 2018). In biofilm producing microorganisms, antimicrobial resistance is much higher than the planktonic cell (Pagedar & Singh, 2012). The prevalence of extended spectrum-β-lactamases (ESBL), Metallo β-lactamase (MBL) and AmpC producing *Pseudomonas aeruginosa* distinguishing virulence factors which may have effects on prolonging treatment and effective control.

The principle of this work was to assess the significant association among the virulence factors, and β-lactamase mechanisms and their antimicrobial resistance of *Pseudomonas aeruginosa*

## 2. MATERIALS AND METHODS

For this study, 120 *P. aeruginosa* isolates of both paediatric and adult patients were obtained from six various types of clinical samples such as blood, wounds, respiratory secretions, stools, urine, and sputum at Selayang Hospital, Kuala Lumpur, Malaysia.

**Antibiotic sensitivity test**

A routine antimicrobial susceptibility test was performed for *Pseudomonas aeruginosa* (120) by Kirby-Bauer disk diffusion standard method. The inoculated bacterial colonies were incubated overnight at 37 ºC. In this study, we have used ten commonly prescribed different types of antibiotics such as tazobactam 10/ piperacillin 75- TZP, ciprofloxacin-CIP, imipenem-IPM, cefepime-FEP, cefotaxime -CTX, aztreonam – ATM, amoxicillin-clavulanic acid – AMC, ceftazidime-CAZ, cefoperazone -CPZ and meropenem-MEM. The results were recorded by measuring the inhibition zone as sensitive, intermediate, and resistant (R), following CLSI guidelines. As a quality control, we have used *P. aeruginosa* ATCC 27853.

**Detection of hemolysin**

For the detection of hemolysins, bacterial isolated colonies were plated on plates (Sheep blood agar 5%) for 24 hours at 37ºC. After incubation, total lysis of red blood cells (clear zone) formed surrounding the cultured colonies which indicated as positivity for hemolysins.

### Detection of phospholipase

The nutrient agar media for detection of phospholipase was inoculated with a bacterial colony for overnight incubation at 37 ºC culture. Formation of bacterial colonies from white to brown color considered as a positive result.

### Detection of gelatinase

The gelatinase production was studied onto nutrient gelatin agar medium containing 1% concentrated bacterial culture. The cultured plates kept for incubation at 37 °C for up to 7 days. Formation of precipitation (an opaque) zone surrounding the spot indicated positivity for proteases.

### Detection of DNAse activity

DNase test agar plates were used in this test to grow the bacterial colonies for 24 to 48 hours at 37°C. HCl/1N solution (one drop) was added to the spotted cultures and a clear zone formation around the culture indicated as positive for DNAse reaction.

### Detection of Biofilm production (tube adherence method)

Tube method is commonly used for biofilm detection. In our study, five bacterial colonies were inoculated to 5ml of BHI broth in glass tubes and kept for incubation at 37°C for 20 hrs. The cultures were suctioned after incubation and saffranine mixed to stains of the test tubes. The visible film growth on the wall of the tube indicated as positivity for biofilm.

Extended-spectrum beta-lactamase (ESBL), AmpC beta-lactamase and Metallo β-lactamase (MBL) genes were detected to find out the antimicrobial mechanisms of *Pseudomonas aeruginosa* by the following methods.

### Detection of Extended spectrum-β-lactamase (ESBL) and Metallo β-lactamase (MBL)

One hundred twenty isolates of *Pseudomonas aeruginosa* were studied to detect the presence of (ESBL and MBL) β-lactamase genes by two most established methods, Double-disk synergy test (DDST) and E-test. In DDST, a cephalosporins (third generation antibiotic) disc and augmentin (penicillin combination) disc were placed 30mm apart from center to center on Muller-Hinton Agar (MHA). A perfect clear zone of inhibition zone around the cephalosporin to augmentin disc was identified for positivity for ESBL gene.

The phenotypic confirmatory test was used among the isolates of *Pseudomonas aeruginosa* for ESBL production. Briefly, a 0.5 MacFarland's suspension of *P. aeruginosa* was spread on a Muller – Hinton agar (MHA) plate and ceftazidime 30 μg (third generation- cephalosporins) disc and ceftazidime 30 μg / clavulanic acid 10 μg (cephalosporins combinations) disc were positioned on the MHA plate. Two discs were placed on the MHA plate with a distance of 15mm in between and kept for overnight incubation at 37°C. Formation of a clear zone of a ≥ 5mm size for the combined discs compared to cefotaxime or ceftazidime disc alone established the presence of ESBL producing an organism. Expansion of zone diameter is a consequence of inhibition of the β-lactamase by clavulanic acid (Thomson, Ayaz, Lutes, & Thomson, 2018). To screen MBL producing *Pseudomonas aeruginosa* strains we have used EDTA, extra pure powder. The inhibition zones of imipenem and imipenem-EDTA disc formed an inhibition zone around them after 24 hours of incubation at 37˚C. The zone of inhibition ≥7mm with imipenem and EDTA disk than imipenem disk alone was detected as MBL producing organisms. Manufacturer's instructions were strictly followed for the E-test for detection of ESBL and MBL producers.

### Detection of AmpC beta-lactamase by disk antagonism test

Disk antagonism test was used to detect inducible AmpC β lactamase among the isolates of *Pseudomonas aeruginosa*. A 0.5 McFarland suspension of bacterial colonies were plated over a Mueller Hinton agar (MHA) plate. Cefotaxime 30μg (cephalosporins- third generation) and cefoxitin 30μg (cephalosporins- second generation) discs were kept 20mm apart from center to center, and the result was observed after overnight incubation at 37°C. AmpC β-lactamase producing isolates were identified by the formation of a blunting inhibition zone of cefotaxime adjacent to cefoxitin (Supriya Upadhyay, Sen, & Bhattacharjee, 2010). AmpC disc test was performed for confirmation of AmpC producing strains.

## 3. RESULTS

One hundred and twenty strains of *Pseudomonas aeruginosa* from six different source (wounds, urine, stools, sputum, respiratory and blood) of clinical samples were used in this study. The percentage of the distribution pattern of *Pseudomonas aeruginosa* in clinical samples is shown in Table 1.

Antimicrobial susceptibility test was done against 10 different types of antibiotic commonly prescribed for the *Pseudomonas aeruginosa* infections. Among the different types of antibiotics, ciprofloxacin showed the highest resistance (38%) followed by ceftazidime (35%) then cefepime (31%) and the lowest percentage of resistance was shown by cefotaxime (10%) (Figure 2).

Forty-nine (40.83%) isolates of *Pseudomonas aeruginosa* showed multidrug resistance. Maximum 10 different types of antibiotics were resistance to 3 (2.5%) isolates followed by 6 antibiotics to 7 (5.8%) and 3 types of antibiotics to 12 (10%) strains of *Pseudomonas aeruginosa* (Table 2).

Virulence factors play an essential role in antimicrobial resistance. In our study, 48.33% isolates were positive for hemolysin followed by 43.33% DNAse, 40.83% phospholipase, 31.66% gelatinase and 17.5% biofilm producer.

Some virulence factors of β-lactamase were observed in 120 clinical strains of *Pseudomonas aeruginosa* in our study. Among all the isolates, 19.16% were ESBL isolates followed by 10.83% AmpC, and 7.5% MBL producers. A total number of 49 (40.83%) isolates were multidrug resistance (Figure 4). The observation of both these tests is shown in figure 5.

An analysis of the association between virulence factors and β-lactamase were performed among all the clinical isolates of *Pseudomonas aeruginosa* (Table 3). We have found that there was a significant association among hemolysin (p-value- 0.02), phospholipase (p-value-0.03), and DNAse (p-value-0.005) in ESBL producers. The MBL producing isolates showed the association with hemolysin (p-value-0.009), gelatinase (p-value-0.0001), and biofilm (p-value-0.02). AmpC producing virulence factors revealed a correlation among gelatinase (p-value-0.00001), DNAse (p-value-0.009), and biofilm (p-value-0.003). We have considered the association to be positive if the p-value is below ($p<0.05$). We have observed the association between virulence factors and β-lactamase producing isolates in our study (Table 3).

## 4. DISCUSSIONS

Severe nosocomial infections are commonly caused by *Pseudomonas aeruginosa*. A leading cause of broad-spectrum infection especially in respiratory, urinary tracts, wounds infections and gastrointestinal is due to *P. aeruginosa*. The pathogenicity of *Pseudomonas aeruginosa* is attributable to the production of multiple virulence factors. Infections caused by *Pseudomonas aeruginosa* is difficult to eliminate because of its poor efficacy of antibiotics and numerous resistance mechanisms by the bacterium (Tran et al., 2014). Some previously published study has reported that the virulence factors are correlated to antibiotic resistance (Roux et al., 2015) and the presence of high level of virulence factors enhance antibacterial resistance (Finlayson & Brown, 2011).

A total of 120 *Pseudomonas aeruginosa* were isolated from six different types of clinical sources (stools, wounds, respiratory, urine, sputum, and blood) and their prevalence was analyzed according to sex, age group, and clinical sources of the samples. Among the clinical isolates, the occurrence rate of *Pseudomonas aeruginosa* was higher at the age group of 31-45 in both male and female. In contrast, the lowest occurance (10%) was observed at the age group of above 60 years (Table 1). According to both sex and age, the differences were shown in the frequency of isolation between males and females.

In this study, Quinolone (antibiotics) showed the highest resistance (Figure 1) which is similar to our another study (Fazlul, Zaini, Rashid, & Nazmul, 2011) followed by cephalosporins groups of antibiotics which are agreeing to another recent study (Senthamarai, 2014). Disk diffusion method was used for antibiotic susceptibility test (Figure 2). Forty-nine (40.83%) isolates were positive for multidrug-resistant while another recent study has shown that 95.8% strains are multidrug-resistant (MDR) (Table 2). The ability of virulence factors significantly associated with multidrug resistance (MDR) (Finlayson & Brown, 2011). It was revealed in some studies that virulence factors and multi-drug resistant (MDR) are significantly associated (Bradbury, Roddam, Merritt, Reid, & Champion, 2010; Deptuła & Gospodarek, 2010; Finlayson & Brown, 2011).

The Pathogenicity of *P. aeruginosa* containing several structural components, toxins, and enzymes play an essential role in forming infectious diseases (Murray, Rosenthal, & Pfaller, 2015). In our study, *Pseudomonas aeruginosa* from different types of clinical samples produces various types of virulence factors. We have observed that the highest 48.33% activity for hemolysin, 43.33% were positive for DNAse, 40.83% positivity for phospholipase, and 31.66% positivity for gelatinase. In contrast, the lowest 17.5% positivity for biofilm production was observed (Figure 3). In our study, the observed virulence factors in *P. aeruginosa* is lower compared to another similar study reported that the positivity for hemolysin was 95.2%, phospholipase 81% and gelatinase 78% ([Khalil, Sonbol, BADR, & ALI, 2015](#)). Interestingly, Pramodhini and her colleague found hemolysin (80.3%), phospholipase (70%) and gelatinase (71.4%) isolates producer ([Pramodhini et al., 2016](#)). Among the virulence factors, phospholipase was detected in 87.5% isolates, 68.75% isolates were biofilm producer, and 81.25% strains were detected as gelatinase positive ([Mohammad, 2013](#)) which were much higher compared to our study.

The lowest percentage of hemolysin (30%) activity was recorded in the blood whereas DNAse (25%) in respiratory. Among all the different sources of clinical samples, urine produced the lowest phospholipase (25%), gelatinase (15%) and biofilm (5%) respectively. The highest occurrence was observed for hemolysin (65%) from stool followed by 60% for phospholipase, gelatinase, DNAse in blood, respiratory, and urine respectively (Figure 3).

*P. aeruginosa* developed its resistance to antibiotics mostly by the occurrence of plasmid-mediated AmpC β-lactamase, extended-spectrum β-lactamases (ESBLs) and Metallo-β-lactamases (MBLs) ([Farshadzadeh, Khosravi, Alavi, Parhizgari, & Hoveizavi, 2014](#)). All the β-lactamases producing strains were revised for the occurrence of several virulence factors. Clinical isolates of ESBL producing *Pseudomonas aeruginosa* under constructive condition express its virulence factors and their association is significant ([Khalil et al., 2015](#); [Mansouri, Norouzi, Moradi, & Nakhaee, 2011](#)). A recent study reported that plasmid-mediated AmpC production in *Pseudomonas aeruginosa* becomes a new risk in the healthcare management of infection ([S Upadhyay, Mishra, Sen, Banerjee, & Bhattacharjee, 2013](#)). Among the 120 isolates of *Pseudomonas aeruginosa*, the highest 19.16% strains were ESBL positive, MBL (7.5%) and 10.83% were AmpC producers (Figure 4). In a recent study, 25.13% isolates were ESBL positive ([Shaikh, Fatima, Shakil, Rizvi, & Kamal, 2015](#)) which is slightly higher compare to our result while 19% ([NagKumar et al., 2015](#)), and 18.37% ([Sachdeva, Sharma, & Sharma, 2017](#)) isolates were MBL producers in *Pseudomonas* which are in agreement with our study. Feglo et al. (2016) ([Feglo & Opoku, 2014](#)) reported 38.0% isolates were AmpC positive while it was 10.83% positivity in our study. The observation of β-lactamase producing *Pseudomonas aeruginosa* was shown in Figure 5. We have observed, virulence factors and β-lactamase producers showed a significant association based on Pearson's correlation coefficient analysis. According to our result, a strong correlation between ESBL producing isolates (19.16%) and virulence factors (hemolysin, phospholipase, and DNAse) was observed. MBL producing strains (7.5%) acted to be more expressive with hemolysin, gelatinase and biofilm and AmpC producing isolates (10.83%) are highly correlated with a phospholipase, gelatinase, DNAse, and biofilm respectively. Standard significance difference was measured by the $p<0.05$ value. No significance was observed among ESBL producing a strain with gelatinase and biofilm factors, MBL producing isolates with phospholipase and DNAse respectively, and AmpC strains with hemolysins (Table 3).

In most of the research, resistance mechanisms and virulence factors have studied separately, but there are a few research which showed the biological association. Pearson's correlation coefficient (($p<0.05$) was used to find out the significant association between virulence factors and β lactamase producers. Varying environmental conditions can enhance the expression of virulence factors in *P. aeruginosa*. Besides, cell signaling pathways regulate the expression of virulence factors ([Ullah et al., 2017](#)). Moreover, effective correlation is present among the secretion of virulence factors and beta-lactamase producing *Pseudomonas aeruginosa* ([Khalil et al., 2015](#)). We could be established that antimicrobial sensitivity of *P. aeruginosa* depends on the microbial biological correlation among virulence factors and source of isolations.

## 5. CONCLUSIONS

This study revealed that the exhibits of virulence factors and the beta-lactamase producers from various infections sites have a strong antimicrobial resistance against different groups of antibiotics and the pathogenesis worsen in healthcare management. From the evaluated data on hemolysin, protease, phospholipase, DNAse, biofilm formation and lactamase among the *Pseudomonas aeruginosa* isolates, an association between the multidrug resistance and pathogenesis are more prominent was observed. Therefore, this study highlights the effect of the correlation between the source of infections and virulence factors must be periodically monitored to understand possible mechanisms. The information and the recommendation of this study will improve to control these infections, development of alternative therapy and therapeutic success in the community.

**Acknowledgments**

Special thanks to all the co-authors for their support to complete this study and there is no conflict of interest to conduct this study.

**Table 1:** Frequency of *Pseudomonas aeruginosa* from different clinical samples collected based on sex, age and sources

| Age Groups in Years | Gender Male | Gender Female | Total (%) | Sources of clinical isolates (120) Stool | Wounds | Respiratory | Urine | Sputum | Blood |
|---|---|---|---|---|---|---|---|---|---|
| 1-15 | 13 | 6 | 15.83 | 5 | 3 | 5 | 2 | 2 | 2 |
| 16-30 | 17 | 9 | 21.67 | 3 | 2 | 5 | 5 | 4 | 7 |
| 31-45 | 25 | 15 | 33.33 | 9 | 10 | 5 | 6 | 6 | 4 |
| 46-60 | 12 | 11 | 19.17 | 2 | 4 | 2 | 6 | 4 | 5 |
| Above 60 | 8 | 4 | 10 | 1 | 1 | 3 | 1 | 4 | 2 |

**Table 2:** Multi-drug resistance among the *Pseudomonas aeruginosa* isolates (120)

| No of antibiotics | Types of antibiotics | No (%) of MDR *P. aeruginosa* isolates |
|---|---|---|
| 3 classes of antibiotics | CTX, TZP, IMP | 12 (10%) |
| 4 classes of antibiotics | ATM, MEM, CIP | 17 (14.1%) |
| 5 classes of antibiotics | MEM, AMC, FEP, CAZ, CTX | 10 (8.3%) |
| 6 classes of antibiotics | AMC, CTX, ATM, IMP, CIP, CPZ | 7 (5.8%) |
| 10 classes of antibiotics | TZP, CAZ, CIP, MEM, IMP, FEP, CTX, ATM, AMC, CPZ | 3 (2.5%) |
| | Total | 49 (40.83%) |

**Table 3:** Analysis of the association between virulence factors and β-lactamase producers

| | | Hemolysin Positive | Hemolysin Negative | Phospholipase Positive | Phospholipase Negative | Gelatinase Positive | Gelatinase Negative | DNAse Positive | DNAse Negative | Biofilm Positive | Biofilm Negative |
|---|---|---|---|---|---|---|---|---|---|---|---|
| Esbl | Positive (23) | 6 | 16 | 5 | 18 | 8 | 15 | 4 | 19 | 6 | 17 |
| | Negative (97) | 52 | 46 | 44 | 53 | 30 | 67 | 48 | 49 | 15 | 82 |
| | P-value | 0.02 | | 0.03 | | 0.07 | | 0.005 | | 0.2 | |
| MBL | Positive (9) | 8 | 1 | 5 | 4 | 8 | 1 | 3 | 6 | 4 | 5 |
| | Negative (113) | 50 | 63 | 44 | 69 | 30 | 83 | 49 | 64 | 17 | 96 |
| | P-value | 0.009 | | 0.3 | | 0.0001 | | 0.5 | | 0.02 | |
| AmpC | Positive (13) | 3 | 10 | 9 | 4 | 11 | 2 | 10 | 3 | 6 | 7 |
| | Negative (107) | 55 | 52 | 40 | 67 | 27 | 80 | 42 | 65 | 15 | 92 |
| | P-value | 0.05 | | 0.02 | | 0.00001 | | 0.009 | | 0.003 | |

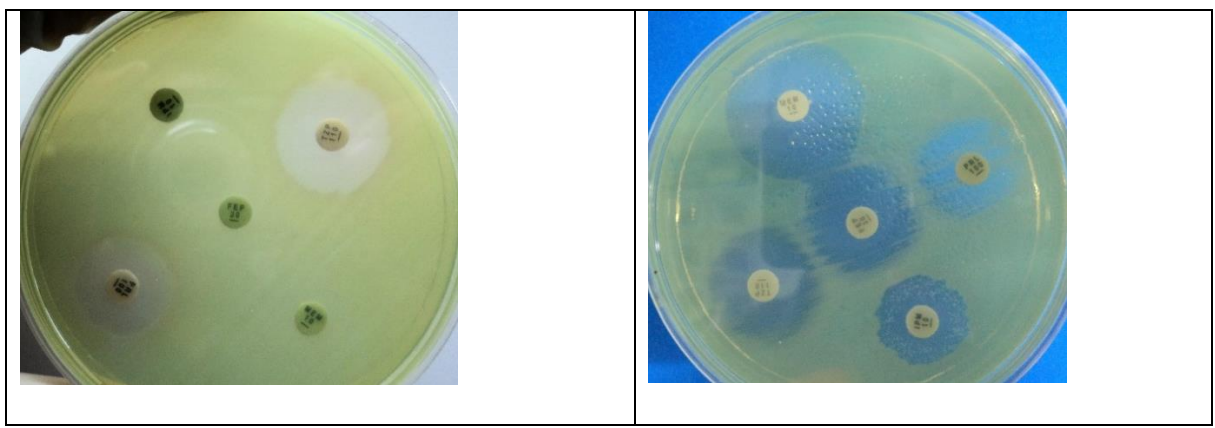

**Figure 1:** Antibiotic susceptibility test by disk diffusion method

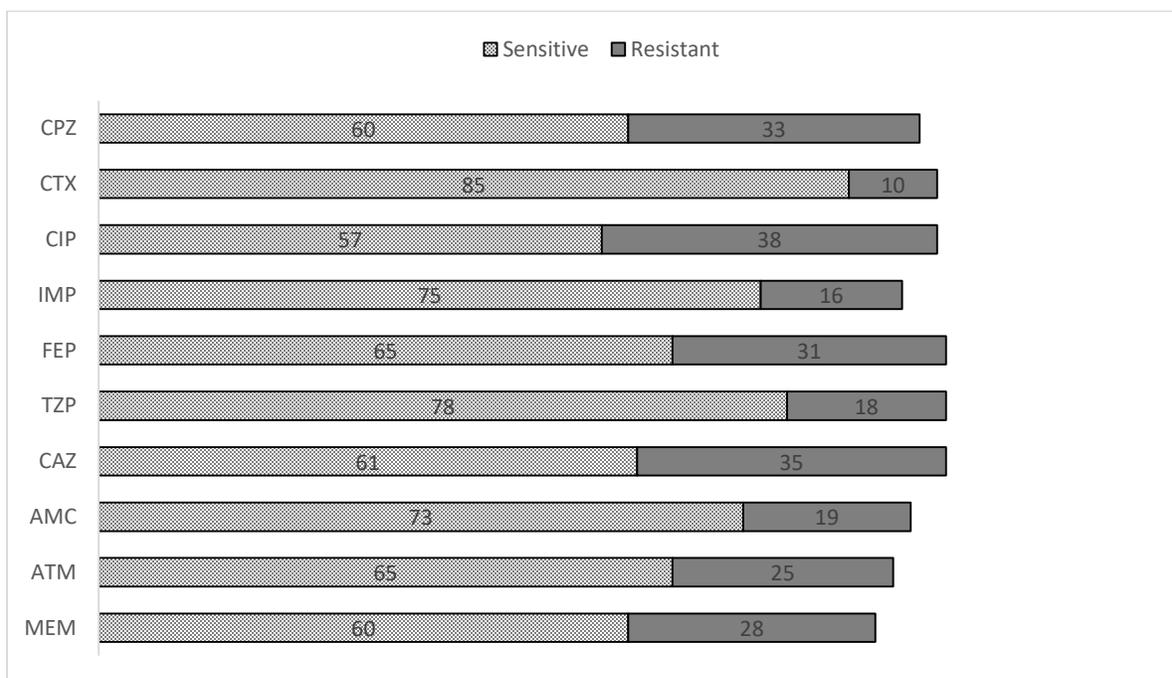

Tazobactam 10/ piperacillin 75- TZP, ciprofloxacin-CIP, imipenem-IPM, cefepime-FEP, cefotaxime -CTX, aztreonam – ATM, amoxicillin-clavulanic acid – AMC, ceftazidime-CAZ, cefoperazone -CPZ and meropenem-MEM.

**Figure 2:** Analysis of antibiotic susceptibility test

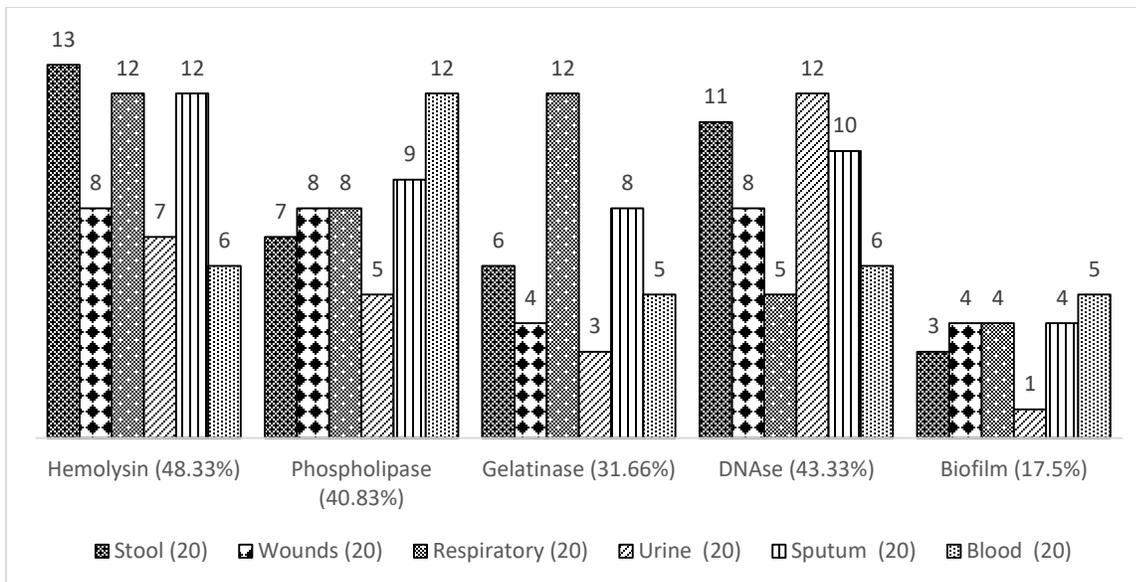

**Figure 3:** Virulence factors production from different clinical sources

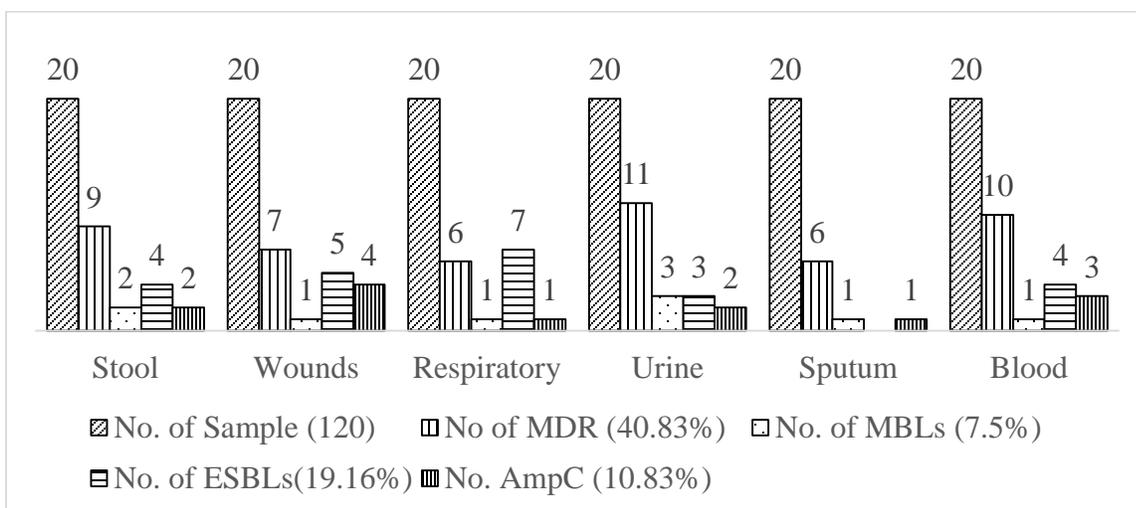

**Figure 4:** ESBL, MBL, AmpC production and multidrug resistance among the isolates (n=120)

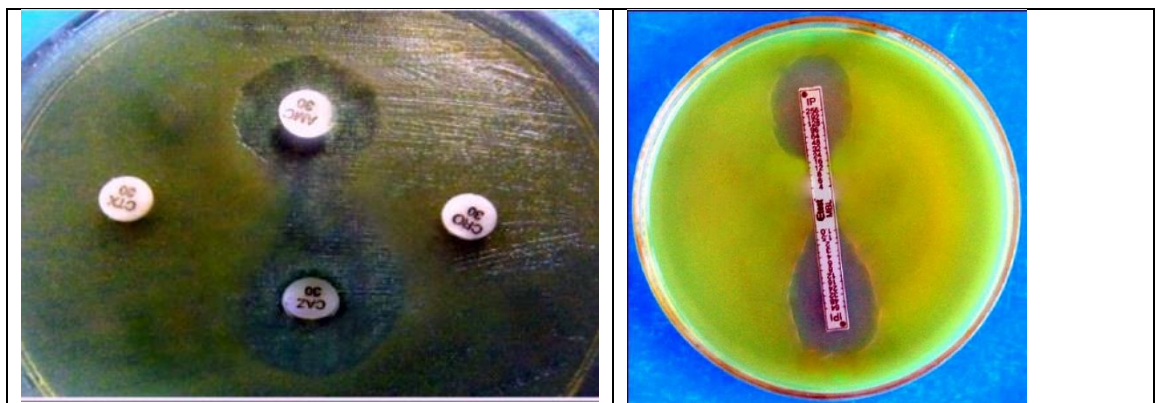

**Figure 5:** Detection of ESBL and MBL test